\documentclass{aastex631}

\begin{document}

\title{BSN: First Photometric Light Curve Analysis of Two W-type Contact Binary Systems\\
OP Boo and V0511 Cam}

\author{Atila Poro}
\altaffiliation{atilaporo@bsnp.info}
\affiliation{Astronomy Department of the Raderon AI Lab., BC., Burnaby, Canada}

\author{Mehmet Tanriver}
\affiliation{Department of Astronomy and Space Science, Faculty of Science, Erciyes University, Kayseri TR-38039, Türkiye}
\affiliation{Erciyes University, Astronomy and Space Science Observatory Application and Research Center, Kayseri TR-38039, Türkiye}

\author{Ahmet Keskin}
\affiliation{Department of Astronomy and Space Science, Faculty of Science, Erciyes University, Kayseri TR-38039, Türkiye}

\author{Ahmet Bulut}
\affiliation{Department of Physics, Faculty of Arts and Sciences, Çanakkale Onsekiz Mart University, Terzioğlu Kampüsü, TR-17020, Çanakkale, Türkiye}
\affiliation{Astrophysics Research Center and Observatory, Çanakkale Onsekiz Mart University, Terzioğlu Kampüsü, TR-17020, Çanakkale, Türkiye}

\author{Salma Rabieefar}
\affiliation{Department of Physics, Khayyam University, Mashhad, Iran}

\author{Malihe Mousapour Gharghabi}
\affiliation{Department of Physics, Khayyam University, Mashhad, Iran}

\author{Filip Walter}
\affiliation{Variable Star and Exoplanet Section of Czech Astronomical Society, Prague, Czech Republic}

\author{Stanislav Holý}
\affiliation{Variable Star and Exoplanet Section of Czech Astronomical Society, Prague, Czech Republic}

\begin{abstract}
This study presented the first light curve analysis of the OP Boo and V0511 Cam binary stars, which was conducted in the frame of the Binary Systems of South and North (BSN) Project. Photometric ground-based observations were conducted with standard filters at two observatories in the Czech Republic. We computed a new ephemeris for each of the systems using our extracted times of minima, TESS data, and additional literature. Linear fits for O-C diagrams of both systems were considered using the Markov Chain Monte Carlo (MCMC) method. The light curves were analyzed using the Wilson-Devinney (WD) binary code combined with the Monte Carlo (MC) simulation. The light curve solutions of both target systems required a cold starspot. The absolute parameters of the systems were calculated by using a $P-M$ parameter relationship. The positions of the systems were also depicted on the Hertzsprung-Russell (HR), $P-L$, $logM_{tot}-logJ_0$, and $T-M$ diagrams. The second component in both systems is determined to be a more massive and hotter star. Therefore, it can be concluded that both systems are W-type contact binary systems.
\end{abstract}


\section{Introduction}
The component stars in a contact binary system overfill their own Roche lobes (\citealt{1968-a-ApJ...151.1123L}). It indicates their surfaces' potentials are equal. The W Ursae Majoris (W UMa) system belongs to a type known as Low-Temperature Contact Binaries (LTCBs), and their stars' temperatures are close to each other (\citealt{rucinski1993contact}, \citealt{2005ApJ...629.1055Y}).
Contact binary systems are divided into two categories, A and W, according to the companions' masses and temperatures (\citealt{binnendijk1970orbital}). The effective temperature of star2 in W-type systems is higher than that of star1.

In addition, the estimation of absolute parameters in contact systems using orbital period has been the goal of many investigations (\citealt{2003MNRAS.342.1260Q}, \citealt{2018PASJ...70...90K}, \citealt{2021ApJS..254...10L}, \citealt{2022MNRAS.510.5315P}). Mass transfer between two stars can also be determined by analyzing variations in the orbital period over time. There are also theories about the upper and lower limits of the orbital period in contact systems, which indicate that it is less than 0.6 days (\citealt{2022MNRAS.510.5315P}).
Although these systems are very important in terms of the formation, stellar structure, and evolution of stars, there are still many ambiguities and questions. It seems that observing and studying more contact systems can help by creating a larger sample to answer the questions (\citealt{2024RAA....24a5002P}).

The first light curve analysis of the OP Boo and V0511 Cam binary systems from the northern hemisphere of the sky was provided in this work. These two binary systems were discovered by the ASAS-SN survey and \cite{2007PZP.....7....6K}. OP Boo and V0511 Cam are introduced as contact binary systems in catalogs and databases.

The OP Boötes (GSC 03861-00642) binary system's coordinates are RA. $225.80046^{\circ}$ and Dec. $53.56501^{\circ}$ from the Gaia\footnote{\url{https://gea.esac.esa.int/archive}} DR3 results. This system's apparent magnitude is reported as $V$=12.78 in the ASAS-SN\footnote{\url{https://asas-sn.osu.edu/variables}} catalog.
OP Boo's orbital period is reported as 0.3114482-day in the ZTF\footnote{The Zwicky Transient Facility (ZTF) catalog of periodic variable stars} catalog, 0.311447-day in the ATLAS catalog, and 0.3114445-day in the ASAS-SN\footnote{\url{https://asas-sn.osu.edu/variables}} catalog.

V0511 Camelopardalis (GSC 04548-01797) is a binary system with coordinates RA. $151.47734^{\circ}$ and Dec. $81.98326^{\circ}$ from Gaia DR3. The ASAS-SN variable stars catalog reported an apparent magnitude of $V$=12.59 for V0511 Cam.
The orbital period of the V0511 Cam system is reported to be 0.4046236-day in the ASAS-SN catalog, and 0.404615-day in the VSX\footnote{AAVSO International Variable Star Index (VSX)} database.

The first light curve study of binary stars is important to create larger samples for deeper parameter investigations of these types of systems. The paper is organized as follows: Specifications on photometric observations and the data reduction process are given in Section 2. For each of the systems, the new ephemeris and extracted minima times are presented in Section 3. The light curve solutions for the systems are contained in Section 4. Section 5 presents the estimation of the absolute parameters. Finally, the conclusion is included in Section 6.

\vspace{1cm}
\section{Observation and Data Reduction}
Two observatories in the Czech Republic observed the binary systems V0511 Cam and OP Boo in an expanse of two nights.

OP Boo was observed using a GSO Newton 200/1000 telescope and a ZWO ASI 178MM CCD. The observation was performed at a private observatory ($49.645^{\circ}$N, $14.755^{\circ}$E) in April 2019.
This observation was carried out with a $V$ filter and a 60-second exposure time. We reduced the raw CCD images, and the basic data reduction was performed for dark and flat-field images using Muniwin 2.1.31 software. During the observation, we used UCAC4 718-055116 ($V^{mag}$=12.95) as a comparison star, UCAC4 718-055147 ($V^{mag}$=12.65) for the first check star, and UCAC4 718-055152 ($V^{mag}$=12.51) for the second check star.

V0511 Cam was observed in September 2021 at the Štefánik Observatory ($50.081^{\circ}$N, $14.398^{\circ}$E). We applied a 16-inch F/10 Schmidt-Cassegrain telescope and a SBIG ST10XME CCD. In this observation, the Johnsons-Cousins $R_c$ filter was used, and the exposure time was 60 seconds. We employed PCs that were online synchronized with stratum 0 NTP servers using Dimension 4 software. The CCD image processing and data reduction were done with dark and flat-field images, relative aperture photometry by Muniwin 2.1 software, and artificial comparison stars from three sources. Therefore, UCAC4 861-006287 ($V^{mag}$=12.26), UCAC4 860-006752 ($V^{mag}$=12.29), and UCAC4 861-006284 ($V^{mag}$=11.97) are used as comparison stars. UCAC4 860-006775 ($V^{mag}$=13.37) was our check star in this observation.

The apparent magnitudes reported in this section for comparison and check stars are from the AAVSO Photometric All Sky Survey (APASS) DR9 catalog.

TESS data were used in this study for both target systems.
TESS observed the OP Boo system in sector 50 with a 600-second exposure. For the V0511 Cam system, we also utilized sector 60 with a 200-second exposure time.
The data is available at the Mikulski Space Telescope Archive (MAST)\footnote{\url{https://mast.stsci.edu/portal/Mashup/Clients/Mast/Portal.htmL}}.

\vspace{1cm}
\section{Orbital Period Variations}
The orbital period of contact systems is known to have changed over time. It is an important parameter obtained from observations to understand some characteristics of these kinds of systems. Orbital period analysis and a new ephemeris computation are important for those systems whose orbital period variations and light curve analysis have not yet been investigated. 

We have extracted a primary and a secondary minimum for each of the OP Boo and V0511 Cam systems. All minima are given in the Barycentric Julian Date and Barycentric Dynamical Time ($BJD_{TDB}$). Table \ref{tab1} contains the times of minima extracted in this study and collected from the literature. Appendix tables \ref{tabA1} and \ref{tabA2} listed the primary and secondary times of minima extracted from TESS data.

For OP Boo, we used a primary minimum (2456191.62078) from the \cite{paschke2014list} study and an orbital period (0.3114445-day) reported by the ASAS-SN catalog as a reference ephemeris. For V0511 Cam, a primary minimum (2451470.67123) from the \cite{2008PZP.....8...52K} study and orbital period (0.4046236-day) come from the ASAS-SN catalog, used for the reference ephemeris. So, the epoch and O-C values of all minima were computed using the reference ephemeris of each system.

\begin{figure*}
\begin{center}
\includegraphics[scale=0.22]{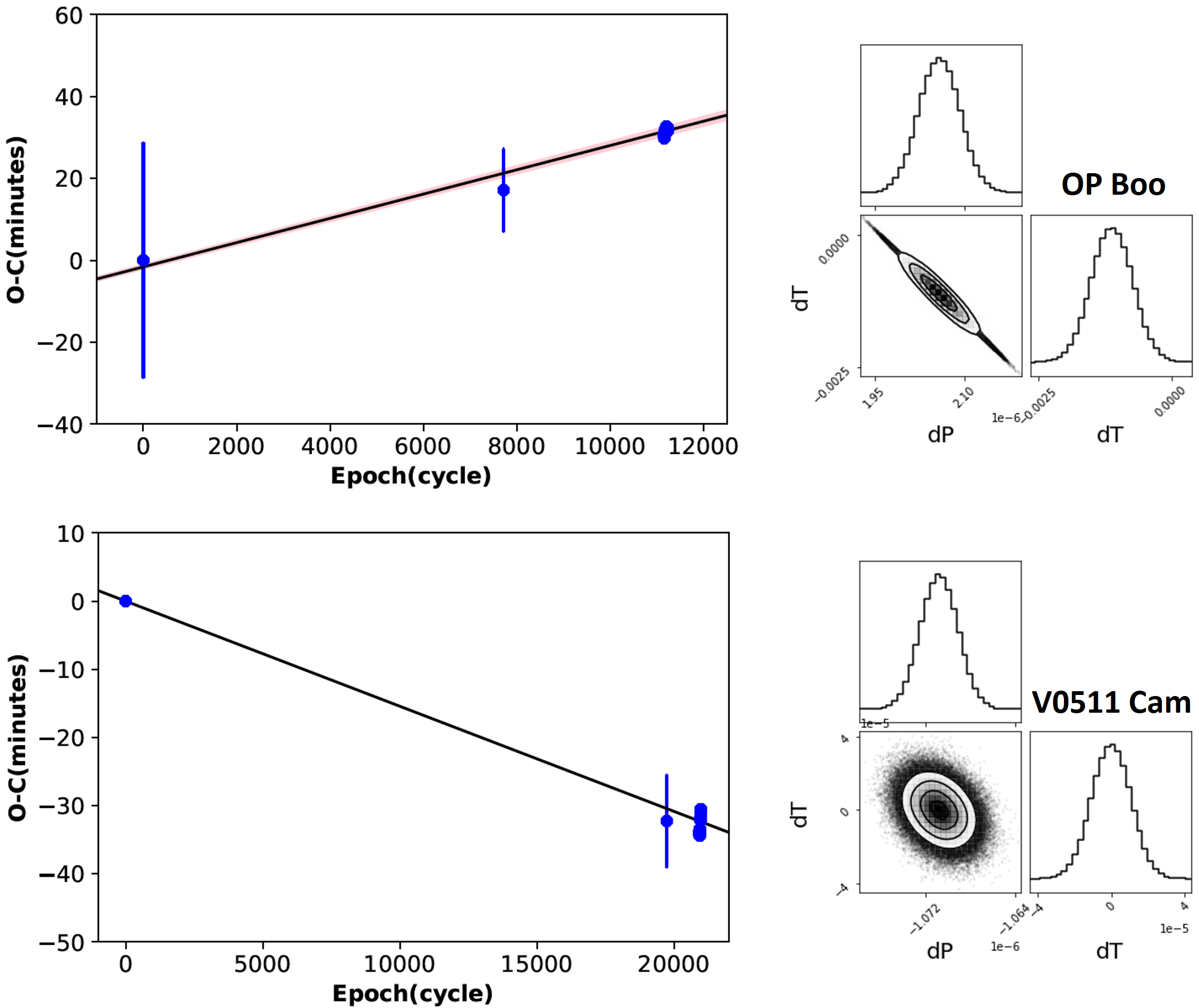}
\caption{The O-C diagrams of OP Boo and V0511 Cam eclipsing binaries with linear fits and corner plots.}
\label{fig1}
\end{center}
\end{figure*}

The number of observed minima and their time intervals are important for O-C analysis. There have been few ground-based observations of the two systems in the study, and linear fits have been taken into consideration for O-C diagrams (Figure \ref{fig1}).
We used 20 walkers and 10000 iterations for each walker in the MCMC process to determine new ephemeris for each system. Thus, we carried out the MCMC sampling using the PyMC3 package (\citealt{salvatier2016pymc3}). The new ephemeris for each system is presented in Equations \ref{eq1} and \ref{eq2}:

\begin{equation}\label{eq1}
OP\ Boo:\ Min.I(BJD_{TDB})=2456191.61897(37)+0.311446559(33)\times E
\end{equation}

\begin{equation}\label{eq2}
V0511\ Cam:\ Min.I(BJD_{TDB})=2451470.67123(1)+0.4046225292(16)\times E
\end{equation}

\begin{table*}
\caption{Ground-based observations' available CCD times of minima.}
\centering
\begin{center}
\footnotesize
\begin{tabular}{c c c c c c c}
 \hline
 \hline
System & Min.($BJD_{TDB}$) & Error & Filter & Epoch & O-C & Reference\\
\hline
OP Boo &2456191.62078	&	0.02000	& CCD &	0 &	0 &	\cite{paschke2014list}	\\
&2457471.49608	&	0.00050	& CCD &	4109.5 & -0.0059 &	\cite{lehky2021brno}	\\
&2458595.36130	& 0.00700 & $V$	& 7718 & 0.0119	& This Study	\\
&2458595.50872	&	0.00660	&	$V$	&	7718.5	& 0.0036	& This Study	\\
\hline
V0511 Cam & 2451470.67123 & & $R$ & 0 & 0 & \cite{2008PZP.....8...52K}\\
& 2456043.51046 & 0.00300 & $-Ir$ & 11301.5 & -0.0144 & \cite{2013IBVS.6048....1H}\\
& 2459459.33503 & 0.00052 & $R_c$ & 19743.5 & -0.0223 & This study\\
& 2459459.53719 & 0.00470 & $R_c$ & 19744 & -0.0224 & This study\\
\hline
\hline
\end{tabular}
\end{center}
\label{tab1}
\end{table*}

\vspace{1cm}
\section{Light Curve Solutions}
Photometric light curve analysis of the OP Boo and V0511 Cam system was performed by the WD code and MC simulation (\citealt{wilson1971realization}).

We assumed the bolometric albedo and gravity-darkening coefficients were $A_1=A_2=0.5$ (\citealt{1969AcA....19..245R}) and $g_1=g_2=0.32$ (\citealt{1967ZA.....65...89L}), respectively. We used the limb darkening coefficients from the \cite{1993AJ....106.2096V} study. Also, we considered the reflection effect in our contact binary systems (\citealt{wilson1990accuracy}, \citealt{2016ApJS..227...29P}).

In this study, we considered the initial system temperature from Gaia. Then, we estimated the components' temperature ratio from the depth difference of the primary and secondary minima. We set the temperature reported from the Gaia DR2 and Gaia DR3 on the hotter stars of OP Boo and V0511 Cam, respectively.

Then, using MC simulation, we performed a mass ratio and other parameters search with large ranges. So, we searched for a mass ratio between 0.1 and 9, inclination between 40 and 90, surface potentials between 1.5 and 9, and temperatures between 4000 and 7000 for both stars (\citealt{2021NewA...8601571P}).

After searching and ensuring a suitable theoretical fit, we did the MC simulations with the five main parameters $i$, $q$, $\Omega$, $T_c$, and $T_h$. It should be noted that the error rate of normalized flux in the TESS data was high (about 0.1), and the analysis was carried out regardless.

According to the asymmetry in the maximum of the light curve, we added a cold starspot on the hotter component of the OP Boo and V0511 Cam systems. This can be described by the O'Connell effect that contact systems are known for their magnetic activity (\citealt{o1951so}).
Figure \ref{fig3} shows that OP Boo's starspot is located near the contact region, which is to find the best synthetic fit on the light curve. Furthermore, the starspot on the V0511 Cam system has a lower temperature than the OP Boo due to the difference in light curve maxima.

\begin{figure*}
\begin{center}
\includegraphics[scale=0.25]{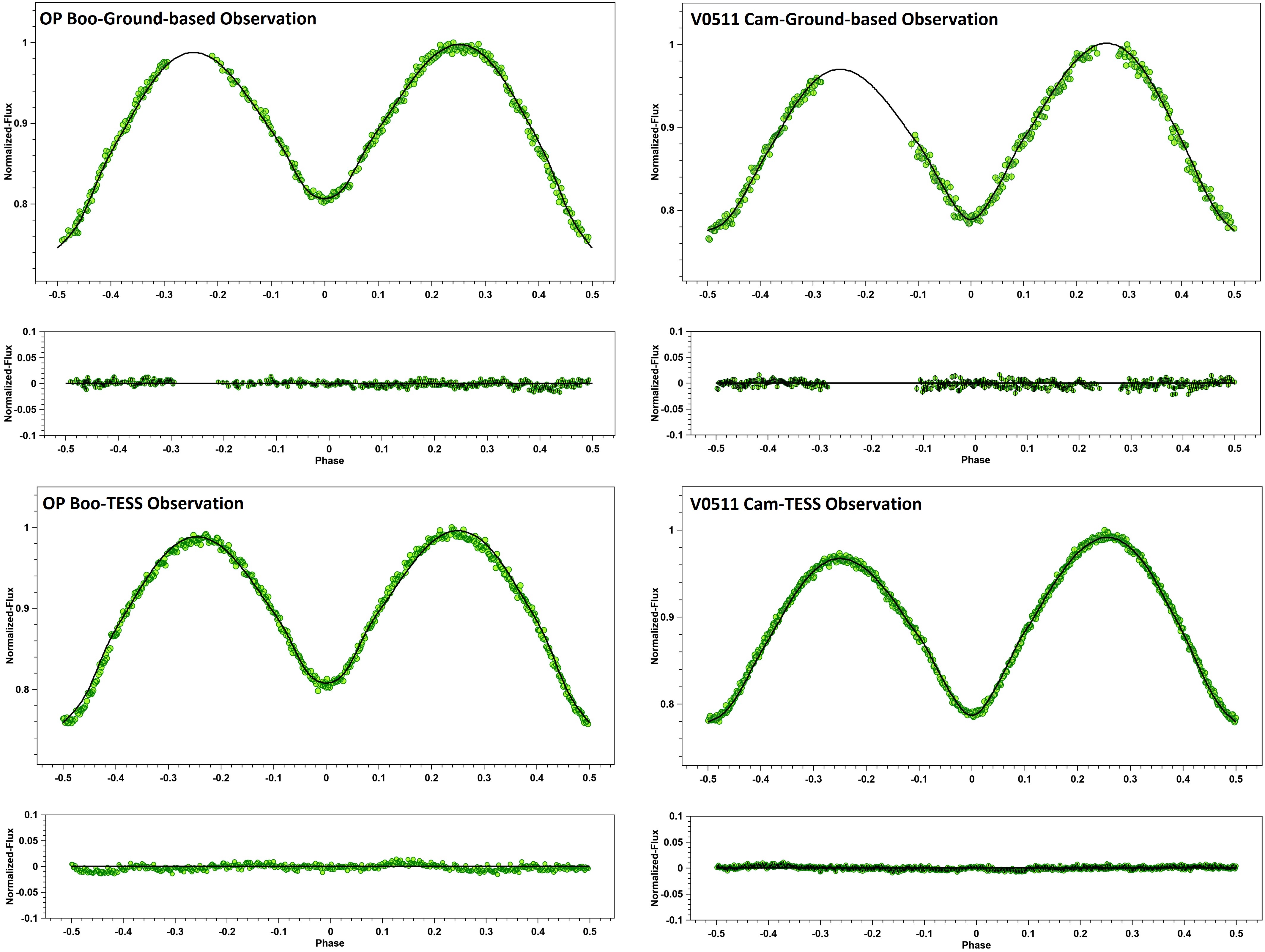}
\caption{The photometric light curves of the systems (green dots), and synthetic light curves obtained from light curve solutions and residuals are plotted.}
\label{fig2}
\end{center}
\end{figure*}

\begin{figure*}
\begin{center}
\includegraphics[scale=0.18]{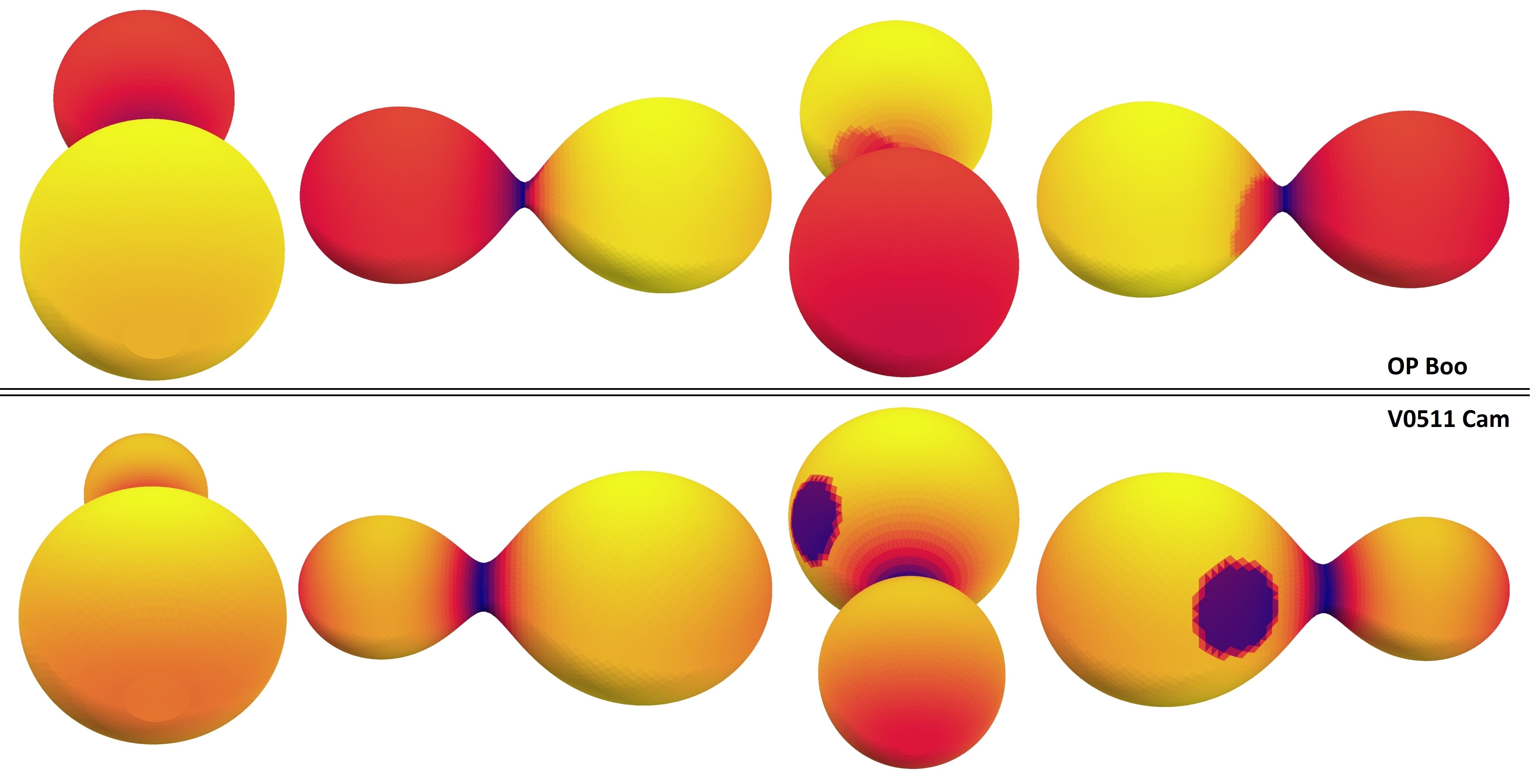}
\caption{3D view of the binary systems in 0, 0.25, 0.5, and 0.75 phases.}
\label{fig3}
\end{center}
\end{figure*}

The results of the light curve analysis are shown in Table \ref{tab2}. The observed and synthetic light curves of the V0511 Cam and OP Boo binary systems are displayed in Figure \ref{fig2}. Furthermore, the geometric structures of the systems are shown in Figures \ref{fig3}.

\begin{table*}
\caption{Photometric solutions of the OP Boo and V0511 Cam systems.}
\centering
\begin{center}
\footnotesize
\begin{tabular}{c c c | c c c}
 \hline
 \hline
Parameter & OP Boo & V0511 Cam & Parameter & OP Boo & V0511 Cam\\
\hline
$T_{c}$ (K) & 4852(27) & 5809(38) & $r_{c(mean)}$ & 0.356(5) & 0.306(5)\\
$T_{h}$ (K) & 5388(32) & 5908(38) & $r_{h(mean)}$ & 0.402(7) & 0.487(7)\\
$q=M_2/M_1$ & 1.233(41) & 2.856(65) & Phase shift & 0.069(1) & -0.056(1)\\
$\Omega_c=\Omega_h$ & 4.100(15) & 6.298(22) & $Col._{spot}$(deg) & 87 & 107\\
$i^{\circ}$ & 59.57(23) & 59.19(39) & $Long._{spot}$(deg) & 344 & 305\\
$f$ & 0.035(4) & 0.206(33) & $Rad._{spot}$(deg) & 21 & 20\\
$l_c/l_{tot}$ & 0.359(1) & 0.278(1) & $T_{spot}/T_{star}$ & 0.95 & 0.90\\
$l_h/l_{tot}$ & 0.641(2) & 0.722(2) & $Component_{spot}$ & Hotter & Hotter \\
\hline
\hline
\end{tabular}
\end{center}
\label{tab2}
\end{table*}

\vspace{1cm}
\section{Estimation of Absolute Parameters}
There is a method to estimate the absolute parameters through Gaia's parallax (\citealt{2022MNRAS.510.5315P}). This method can be accurate and if the values of $a_1(R_{\odot})$ and $a_2(R_{\odot})$ are close together, it is also possible to conclude that the light curve solution makes sense (\citealt{2023RAA....23i5011P}). However, the possibility of using this method is dependent on the extinction coefficient $V_{max}$ from the observation and the appropriate $A_V$. The precision of computations, including parallax, is reduced with increasing values of $A_V$ (\citealt{2024PASP..136b4201P}).
Regarding the OP Boo binary system, in Gaia DR3 the value of the parallax error (0.5880) exceeds that of the parallax (0.4216) and this causes a large error in the distance value ($d(pc)=2371.99\pm 990.48$). However, we can also look at the Re-normalized Unit Weight Error (RUWE) in Gaia DR3, whose value for this system is 54.94 but should be less than 1.4 (\citealt{lindegren2018re}).
On the other hand, values of $d(pc)=634.17\pm 4.35$ and RUWE=1.02 are acceptable and appropriate for the V0511 Cam system. So, Gaia DR3 parallax is not a good way to estimate the absolute parameters of the OP Boo system.

There are other methods for estimating absolute parameters, most of which use sample-based statistical analysis. So, we employed the $P-M_{1}$ relation from the \cite{2022MNRAS.510.5315P} study (Equation \ref{eq3}). This relation is related to a more massive component.

\begin{equation}\label{eq3}
M_{1}=(2.924\pm0.075)P+(0.147\pm0.029)
\end{equation}

This relation comes from a sample of 118 systems, and for all of them, the Gaia parallax was used to estimate absolute parameters. Therefore, its output may closely resemble that of the Gaia Parallax used directly.

First, we estimated the mass of the more massive star using the $P-M_1$ relation. The mass ratio is then employed to determine the mass of the other star. $a(R_{\odot})$ was calculated using the system's total mass and orbital period, and the radius of each star can be found using $r_{mean}$.
Additionally, the stars' luminosity was determined using each star's radius and temperature.
Finally, we calculated the absolute bolometric ($M_{bol}$) of the stars using the well-known Pogson’s relation (\citealt{1856MNRAS..17...12P}), where $M_{bol\odot}$ is taken as 4.73-mag.
The estimation of absolute parameters of OP Boo and V0511 Cam systems is presented in Table \ref{tab3}.

\begin{table*}
\caption{The absolute parameters of OP Boo and V0511 Cam.}
\centering
\begin{center}
\footnotesize
\begin{tabular}{c c c c c c c}
 \hline
 \hline
&& \multicolumn{2}{c}{OP Boo} && \multicolumn{2}{c}{V0511 Cam}\\
Parameter && Cooler star & Hotter star && Cooler star & Hotter star\\
\hline
$M(M_\odot)$ && 0.858(13) & 1.058(52) && 0.463(12) & 1.323(66)\\
$R(R_\odot)$ && 0.855(84) & 0.965(99) && 0.854(107) & 1.360(167)\\
$L(L_\odot)$ && 0.365(85) & 0.708(173) && 0.750(224) & 2.032(595)\\
$M_{bol}(mag.)$ && 5.824(228) & 5.105(237) && 5.043(284) & 3.960(279)\\
$log(g)(cgs)$ && 4.508(75) & 4.493(64) && 4.240(91) & 4.292(79)\\
$a(R_\odot)$ && \multicolumn{2}{c}{2.401(200)} && \multicolumn{2}{c}{2.793(298)}\\
\hline
\hline
\end{tabular}
\end{center}
\label{tab3}
\end{table*}

\vspace{1cm}
\section{Conclusion}
Photometric observations of the OP Boo and V0511 Cam systems were carried out at two observatories in the Czech Republic. Data reduction processes were done according to the standard method, and light curves were prepared for analysis. We also used TESS data for both binary systems.

We extracted the times of minima from our observations and TESS data. Then, we collected mid-eclipse times from the literature as well. Using the reference ephemeris, the epoch and O-C values were calculated. We used the MCMC method for linear fits in the O-C diagrams and presented a new ephemeris for each system.

We presented the first light curve analysis for both target binary systems in this study. Light curve analysis was done with the WD code and MC simulation. The temperature obtained from the light curve solutions shows that in both systems, the secondary minimum is deeper and star 2 is hotter than star 1.
The temperature difference between the two stars in the OP Boo system is 536 K, and in the V0511 Cam system, it is 99 K.
According to the temperature of each star obtained from the light curve analysis and the \cite{2000asqu.book.....C} study, it is possible to determine their spectral type.
Therefore, for the OP Boo system, the cooler star is K2 and the hotter star is K0 in the spectral type; this is for V0511 Cam's cooler star which is G3, and G2 is for the hotter star.
The solution of the light curve for the OP Boo and V0511 Cam systems required the addition of a cold starspot on the hotter component, representing the O'Connell effect (\citealt{o1951so}).

\begin{figure*}
\begin{center}
\includegraphics[width=\textwidth]{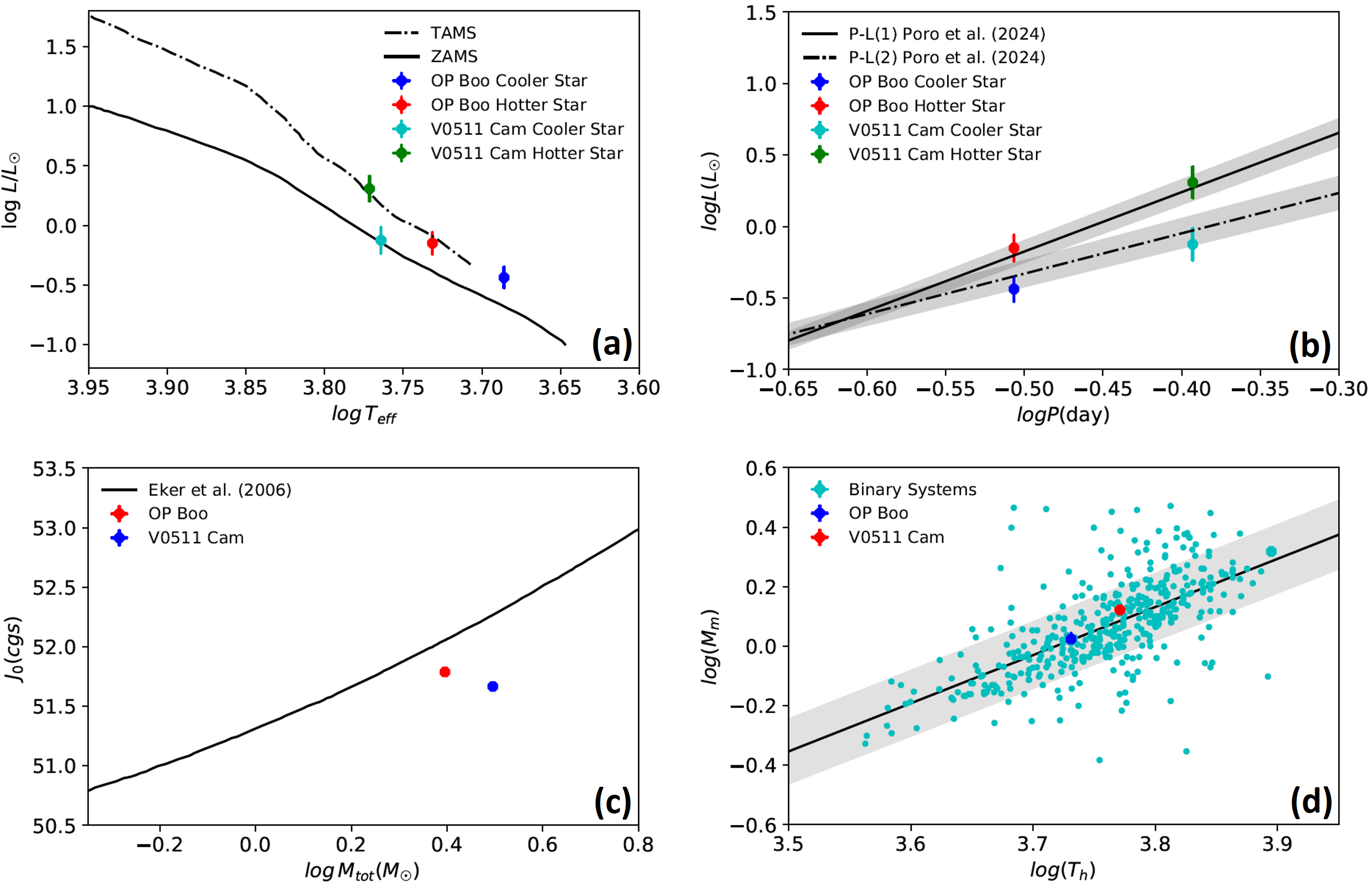}
\caption{(a) HR, (b) $P-L$, and (c) $logM_{tot}-logJ_0$ (d) $T_h-M_m$ diagrams, respectively.}
\label{fig4}
\end{center}
\end{figure*}

We estimated the absolute parameters of the systems using the relationship between the orbital period and the mass of the more massive component (\citealt{2022MNRAS.510.5315P}) and the results of the light curve analysis.
The positions of OP Boo and V0511 Cam stars on the HR diagram are presented (Figure \ref{fig4}a). So, the HR diagram shows the hotter stars are on the Terminal-Age Main Sequence (TAMS) line, and the cooler components lie between the Zero-Age Main Sequence (ZAMS) and TAMS. Figure \ref{fig4}b shows the position of the stars of the two systems compared to the $P-L_{1,2}$ theoretical fit obtained from the \cite{2024RAA....24a5002P} study, which is in good agreement.
As expected, hotter and more massive stars are on the $P-L_1$ theoretical fit, and cooler and less massive stars are on the $P-L_2$ theoretical fit.

Based on computations, the orbital angular momentum of OP Boo is $51.778\pm0.023$, and the value of V0511 Cam is $51.666\pm0.026$. So, OP Boo and V0511 Cam are located in a contact binary systems region, as shown by the $logM_{tot}-logJ_0$ diagram (Figure \ref{fig4}c). The parabolic curve is shown in Figure \ref{fig4}c, and the results are based on the \cite{2006MNRAS.373.1483E} study.

The \cite{2024RAA....24e5001P} study used 428 contact binary systems and presented the $T_h-M_m$ relationship. The position of the target systems is shown in this relationship. In Figure \ref{fig4}d, the horizontal axis shows the effective temperature of the hotter component, and the vertical axis shows the more massive star of each system. As Figure \ref{fig4}d shows, the positions of the stars are in good agreement with the theoretical fit.

We obtained a mass ratio, fillout factor, inclination, and stars' temperature from the light curve solution and MC simulation. The light curve analysis and absolute parameters of both systems suggest that OP Boo and V0511 Cam are W UMa contact and W-subtype binary systems.

\vspace{1cm}
\section*{Data availability}
Data will be made available on request.

\vspace{1cm}
\section*{Acknowledgements}
This manuscript has been prepared based on a multilateral collaboration between the BSN project, the Raderon AI Lab (\url{https://raderonlab.ca}), and Erciyes University (\url{https://www.erciyes.edu.tr}). "This study was supported by the Scientific Research Projects Coordination Unit of Erciyes University (project number FBA-2022-11737)". The authors thank the Astronomical Society of the Czech Republic; data is available in \url{http://var2.astro.cz/}. We made use of information from the Gaia mission of the European Space Agency\footnote{\url{http://www.cosmos.esa.int/gaia}}. Data from the TESS mission observations are used in this paper. The NASA Explorer Program provides funding for the TESS project. We are grateful to Ehsan Paki and Somayeh Soomandar for their help.

\vspace{1cm}
\section*{ORCID iDs}
\noindent Atila Poro: 0000-0002-0196-9732\\
Mehmet Tanriver: 0000-0002-3263-9680\\
Ahmet Keskin: 0000-0002-9314-0648\\
Ahmet Bulut: 0000-0002-7215-926X\\
Salma Rabieefar: 0009-0001-9938-9970\\
Malihe Mousapour Gharghabi: 0009-0000-5748-8316\\
Filip Walter: 0000-0003-2060-4912\\

\vspace{1cm}
\section*{Appendix}
The appendix tables contain a list of the primary and secondary times of minima for the OP Boo and V0511 Cam systems, extracted from TESS observations.

\begin{table*}
\caption{The extracted primary times of minima from TESS sector 50 data for OP Boo.}
\centering
\begin{center}
\footnotesize
\begin{tabular}{c c c c c c c c c c c c}
\hline
\hline
Min. & Error & Epoch & O-C & Min. & Error & Epoch & O-C & Min. & Error & Epoch & O-C\\
\hline
2459665.3400	&	0.0002	&	11153.5	&	0.0230	&	2459673.1259	&	0.0002	&	11178.5	&	0.0227	&	2459681.6901	&	0.0002	&	11206	&	0.0223	\\
2459665.4938	&	0.0002	&	11154	&	0.0211	&	2459673.2810	&	0.0001	&	11179	&	0.0222	&	2459681.8469	&	0.0003	&	11206.5	&	0.0233	\\
2459665.6513	&	0.0002	&	11154.5	&	0.0228	&	2459673.4372	&	0.0002	&	11179.5	&	0.0227	&	2459682.0013	&	0.0003	&	11207	&	0.0220	\\
2459665.8055	&	0.0002	&	11155	&	0.0213	&	2459673.5922	&	0.0002	&	11180	&	0.0219	&	2459682.1582	&	0.0003	&	11207.5	&	0.0232	\\
2459665.9631	&	0.0002	&	11155.5	&	0.0232	&	2459673.7488	&	0.0002	&	11180.5	&	0.0228	&	2459682.3131	&	0.0003	&	11208	&	0.0224	\\
2459666.1163	&	0.0002	&	11156	&	0.0207	&	2459673.9036	&	0.0002	&	11181	&	0.0219	&	2459682.4697	&	0.0002	&	11208.5	&	0.0232	\\
2459666.2748	&	0.0002	&	11156.5	&	0.0234	&	2459674.0604	&	0.0002	&	11181.5	&	0.0229	&	2459682.6245	&	0.0002	&	11209	&	0.0223	\\
2459666.4280	&	0.0002	&	11157	&	0.0210	&	2459674.2150	&	0.0001	&	11182	&	0.0219	&	2459682.7809	&	0.0002	&	11209.5	&	0.0230	\\
2459666.5859	&	0.0002	&	11157.5	&	0.0231	&	2459674.3716	&	0.0001	&	11182.5	&	0.0227	&	2459682.9359	&	0.0003	&	11210	&	0.0222	\\
2459666.7394	&	0.0003	&	11158	&	0.0209	&	2459674.5267	&	0.0002	&	11183	&	0.0220	&	2459683.0925	&	0.0002	&	11210.5	&	0.0232	\\
2459666.8971	&	0.0003	&	11158.5	&	0.0228	&	2459674.6832	&	0.0001	&	11183.5	&	0.0228	&	2459683.2475	&	0.0003	&	11211	&	0.0224	\\
2459667.0513	&	0.0003	&	11159	&	0.0214	&	2459674.8380	&	0.0001	&	11184	&	0.0220	&	2459683.4040	&	0.0001	&	11211.5	&	0.0232	\\
2459667.2085	&	0.0002	&	11159.5	&	0.0228	&	2459674.9948	&	0.0001	&	11184.5	&	0.0230	&	2459683.5588	&	0.0002	&	11212	&	0.0223	\\
2459667.3628	&	0.0002	&	11160	&	0.0214	&	2459675.1495	&	0.0002	&	11185	&	0.0220	&	2459683.7156	&	0.0001	&	11212.5	&	0.0234	\\
2459667.5201	&	0.0003	&	11160.5	&	0.0230	&	2459675.3062	&	0.0001	&	11185.5	&	0.0230	&	2459683.8703	&	0.0001	&	11213	&	0.0224	\\
2459667.6745	&	0.0003	&	11161	&	0.0216	&	2459675.4606	&	0.0002	&	11186	&	0.0216	&	2459684.0269	&	0.0003	&	11213.5	&	0.0233	\\
2459667.8316	&	0.0003	&	11161.5	&	0.0230	&	2459675.6178	&	0.0002	&	11186.5	&	0.0231	&	2459684.1819	&	0.0001	&	11214	&	0.0225	\\
2459667.9857	&	0.0002	&	11162	&	0.0214	&	2459675.7722	&	0.0001	&	11187	&	0.0218	&	2459684.3384	&	0.0001	&	11214.5	&	0.0233	\\
2459668.1432	&	0.0002	&	11162.5	&	0.0232	&	2459675.9291	&	0.0002	&	11187.5	&	0.0230	&	2459684.4930	&	0.0002	&	11215	&	0.0222	\\
2459668.2973	&	0.0003	&	11163	&	0.0216	&	2459676.0834	&	0.0001	&	11188	&	0.0216	&	2459684.6497	&	0.0001	&	11215.5	&	0.0232	\\
2459668.4541	&	0.0002	&	11163.5	&	0.0226	&	2459676.2405	&	0.0001	&	11188.5	&	0.0230	&	2459684.8046	&	0.0001	&	11216	&	0.0223	\\
2459668.6080	&	0.0003	&	11164	&	0.0208	&	2459676.3950	&	0.0001	&	11189	&	0.0217	&	2459684.9612	&	0.0001	&	11216.5	&	0.0232	\\
2459668.7659	&	0.0002	&	11164.5	&	0.0230	&	2459676.5522	&	0.0002	&	11189.5	&	0.0232	&	2459685.1162	&	0.0002	&	11217	&	0.0225	\\
2459668.9199	&	0.0002	&	11165	&	0.0212	&	2459676.7064	&	0.0001	&	11190	&	0.0217	&	2459685.2727	&	0.0002	&	11217.5	&	0.0233	\\
2459669.0771	&	0.0001	&	11165.5	&	0.0228	&	2459676.8634	&	0.0001	&	11190.5	&	0.0230	&	2459685.4275	&	0.0002	&	11218	&	0.0223	\\
2459669.2314	&	0.0003	&	11166	&	0.0213	&	2459677.0180	&	0.0002	&	11191	&	0.0218	&	2459685.5841	&	0.0002	&	11218.5	&	0.0232	\\
2459669.3883	&	0.0001	&	11166.5	&	0.0225	&	2459677.1749	&	0.0002	&	11191.5	&	0.0230	&	2459685.7390	&	0.0002	&	11219	&	0.0224	\\
2459669.5429	&	0.0002	&	11167	&	0.0214	&	2459677.3294	&	0.0001	&	11192	&	0.0218	&	2459685.8956	&	0.0002	&	11219.5	&	0.0233	\\
2459669.6998	&	0.0002	&	11167.5	&	0.0226	&	2459677.4863	&	0.0001	&	11192.5	&	0.0229	&	2459686.0503	&	0.0001	&	11220	&	0.0223	\\
2459669.8543	&	0.0003	&	11168	&	0.0213	&	2459677.6408	&	0.0001	&	11193	&	0.0217	&	2459686.2072	&	0.0001	&	11220.5	&	0.0234	\\
2459670.0113	&	0.0002	&	11168.5	&	0.0226	&	2459677.7979	&	0.0002	&	11193.5	&	0.0231	&	2459686.3618	&	0.0001	&	11221	&	0.0222	\\
2459670.1660	&	0.0001	&	11169	&	0.0216	&	2459677.9527	&	0.0001	&	11194	&	0.0221	&	2459686.5183	&	0.0002	&	11221.5	&	0.0231	\\
2459670.3227	&	0.0001	&	11169.5	&	0.0226	&	2459678.1095	&	0.0002	&	11194.5	&	0.0233	&	2459686.6731	&	0.0001	&	11222	&	0.0222	\\
2459670.4773	&	0.0002	&	11170	&	0.0215	&	2459678.2638	&	0.0001	&	11195	&	0.0219	&	2459686.8298	&	0.0001	&	11222.5	&	0.0231	\\
2459670.6341	&	0.0002	&	11170.5	&	0.0226	&	2459679.3550	&	0.0008	&	11198.5	&	0.0230	&	2459686.9848	&	0.0002	&	11223	&	0.0224	\\
2459670.7890	&	0.0002	&	11171	&	0.0217	&	2459679.5098	&	0.0002	&	11199	&	0.0221	&	2459687.1412	&	0.0002	&	11223.5	&	0.0230	\\
2459670.9454	&	0.0002	&	11171.5	&	0.0224	&	2459679.6673	&	0.0002	&	11199.5	&	0.0239	&	2459687.2962	&	0.0002	&	11224	&	0.0224	\\
2459671.1003	&	0.0002	&	11172	&	0.0215	&	2459679.8213	&	0.0003	&	11200	&	0.0222	&	2459687.4528	&	0.0001	&	11224.5	&	0.0232	\\
2459671.2570	&	0.0002	&	11172.5	&	0.0226	&	2459679.9781	&	0.0002	&	11200.5	&	0.0232	&	2459687.6074	&	0.0002	&	11225	&	0.0221	\\
2459671.4119	&	0.0001	&	11173	&	0.0217	&	2459680.1331	&	0.0002	&	11201	&	0.0224	&	2459687.7642	&	0.0001	&	11225.5	&	0.0231	\\
2459671.5686	&	0.0003	&	11173.5	&	0.0227	&	2459680.2894	&	0.0003	&	11201.5	&	0.0230	&	2459687.9186	&	0.0002	&	11226	&	0.0219	\\
2459671.7233	&	0.0001	&	11174	&	0.0217	&	2459680.4440	&	0.0003	&	11202	&	0.0220	&	2459688.0756	&	0.0002	&	11226.5	&	0.0232	\\
2459671.8800	&	0.0002	&	11174.5	&	0.0227	&	2459680.6011	&	0.0002	&	11202.5	&	0.0233	&	2459688.2305	&	0.0002	&	11227	&	0.0223	\\
2459672.0347	&	0.0002	&	11175	&	0.0217	&	2459680.7562	&	0.0002	&	11203	&	0.0227	&	2459688.3873	&	0.0002	&	11227.5	&	0.0234	\\
2459672.1915	&	0.0001	&	11175.5	&	0.0227	&	2459680.9126	&	0.0002	&	11203.5	&	0.0233	&	2459688.5417	&	0.0001	&	11228	&	0.0221	\\
2459672.5029	&	0.0002	&	11176.5	&	0.0227	&	2459681.0673	&	0.0002	&	11204	&	0.0224	&	2459688.6987	&	0.0002	&	11228.5	&	0.0233	\\
2459672.6579	&	0.0002	&	11177	&	0.0220	&	2459681.2238	&	0.0002	&	11204.5	&	0.0231	&	2459688.8531	&	0.0001	&	11229	&	0.0220	\\
2459672.8144	&	0.0002	&	11177.5	&	0.0227	&	2459681.3788	&	0.0002	&	11205	&	0.0224	&	2459689.0100	&	0.0002	&	11229.5	&	0.0232	\\
2459672.9695	&	0.0002	&	11178	&	0.0221	&	2459681.5354	&	0.0002	&	11205.5	&	0.0233	&	2459689.1650	&	0.0002	&	11230	&	0.0225	\\
\hline
\hline
\end{tabular}
\end{center}
\label{tabA1}
\end{table*}

\begin{table*}
\caption{The extracted primary times of minima from TESS sector 60 data for V0511 Cam.}
\centering
\begin{center}
\footnotesize
\begin{tabular}{c c c c c c c c c c c c}
\hline
\hline
Min. & Error & Epoch & O-C & Min. & Error & Epoch & O-C\\
\hline
2459939.6259	&	0.0002	&	20930.5	&	-0.0196	&	2459948.5276	&	0.0002	&	20952.5	&	-0.0196	\\
2459939.8241	&	0.0003	&	20931	&	-0.0237	&	2459948.7260	&	0.0003	&	20953	&	-0.0235	\\
2459940.0310	&	0.0002	&	20931.5	&	-0.0191	&	2459948.9321	&	0.0002	&	20953.5	&	-0.0197	\\
2459940.2289	&	0.0002	&	20932	&	-0.0235	&	2459949.1306	&	0.0003	&	20954	&	-0.0235	\\
2459940.4357	&	0.0002	&	20932.5	&	-0.0191	&	2459949.3371	&	0.0002	&	20954.5	&	-0.0194	\\
2459940.6335	&	0.0002	&	20933	&	-0.0235	&	2459949.5353	&	0.0002	&	20955	&	-0.0235	\\
2459940.8402	&	0.0002	&	20933.5	&	-0.0191	&	2459949.7404	&	0.0002	&	20955.5	&	-0.0207	\\
2459941.0380	&	0.0002	&	20934	&	-0.0237	&	2459955.2014	&	0.0002	&	20969	&	-0.0221	\\
2459941.2451	&	0.0002	&	20934.5	&	-0.0189	&	2459955.4054	&	0.0002	&	20969.5	&	-0.0204	\\
2459941.4426	&	0.0003	&	20935	&	-0.0236	&	2459955.6058	&	0.0002	&	20970	&	-0.0223	\\
2459941.6498	&	0.0002	&	20935.5	&	-0.0189	&	2459955.8098	&	0.0001	&	20970.5	&	-0.0206	\\
2459941.8473	&	0.0003	&	20936	&	-0.0236	&	2459956.0106	&	0.0002	&	20971	&	-0.0222	\\
2459942.0543	&	0.0002	&	20936.5	&	-0.0190	&	2459956.2137	&	0.0002	&	20971.5	&	-0.0214	\\
2459942.2520	&	0.0002	&	20937	&	-0.0236	&	2459956.6185	&	0.0002	&	20972.5	&	-0.0212	\\
2459942.4587	&	0.0003	&	20937.5	&	-0.0191	&	2459956.8197	&	0.0002	&	20973	&	-0.0223	\\
2459942.6565	&	0.0002	&	20938	&	-0.0236	&	2459957.0231	&	0.0002	&	20973.5	&	-0.0212	\\
2459942.8635	&	0.0002	&	20938.5	&	-0.0190	&	2459957.2245	&	0.0002	&	20974	&	-0.0221	\\
2459943.0611	&	0.0003	&	20939	&	-0.0237	&	2459957.4275	&	0.0002	&	20974.5	&	-0.0214	\\
2459943.2679	&	0.0002	&	20939.5	&	-0.0192	&	2459957.6290	&	0.0002	&	20975	&	-0.0222	\\
2459943.4659	&	0.0003	&	20940	&	-0.0235	&	2459957.8326	&	0.0002	&	20975.5	&	-0.0209	\\
2459943.6718	&	0.0002	&	20940.5	&	-0.0199	&	2459958.0340	&	0.0002	&	20976	&	-0.0219	\\
2459944.0775	&	0.0002	&	20941.5	&	-0.0188	&	2459958.2369	&	0.0001	&	20976.5	&	-0.0212	\\
2459944.2748	&	0.0003	&	20942	&	-0.0239	&	2459958.4389	&	0.0002	&	20977	&	-0.0216	\\
2459944.4821	&	0.0002	&	20942.5	&	-0.0189	&	2459958.6411	&	0.0001	&	20977.5	&	-0.0217	\\
2459944.6796	&	0.0002	&	20943	&	-0.0237	&	2459958.8434	&	0.0002	&	20978	&	-0.0217	\\
2459944.8867	&	0.0002	&	20943.5	&	-0.0189	&	2459959.0459	&	0.0001	&	20978.5	&	-0.0215	\\
2459945.0841	&	0.0002	&	20944	&	-0.0238	&	2459959.2482	&	0.0002	&	20979	&	-0.0215	\\
2459945.2912	&	0.0002	&	20944.5	&	-0.0190	&	2459959.4502	&	0.0002	&	20979.5	&	-0.0218	\\
2459945.4888	&	0.0003	&	20945	&	-0.0237	&	2459959.6527	&	0.0002	&	20980	&	-0.0217	\\
2459945.6958	&	0.0002	&	20945.5	&	-0.0190	&	2459959.8548	&	0.0001	&	20980.5	&	-0.0219	\\
2459945.8937	&	0.0002	&	20946	&	-0.0235	&	2459960.0578	&	0.0002	&	20981	&	-0.0212	\\
2459946.1001	&	0.0002	&	20946.5	&	-0.0193	&	2459960.2595	&	0.0001	&	20981.5	&	-0.0218	\\
2459946.2985	&	0.0002	&	20947	&	-0.0232	&	2459960.4621	&	0.0002	&	20982	&	-0.0215	\\
2459946.5051	&	0.0002	&	20947.5	&	-0.0190	&	2459960.6638	&	0.0001	&	20982.5	&	-0.0221	\\
2459946.7029	&	0.0002	&	20948	&	-0.0235	&	2459960.8669	&	0.0002	&	20983	&	-0.0214	\\
2459946.9095	&	0.0002	&	20948.5	&	-0.0192	&	2459961.0687	&	0.0002	&	20983.5	&	-0.0219	\\
2459947.1076	&	0.0003	&	20949	&	-0.0234	&	2459961.2716	&	0.0001	&	20984	&	-0.0213	\\
2459947.3139	&	0.0002	&	20949.5	&	-0.0194	&	2459961.4730	&	0.0001	&	20984.5	&	-0.0221	\\
2459947.5123	&	0.0002	&	20950	&	-0.0234	&	2459961.6763	&	0.0002	&	20985	&	-0.0212	\\
2459947.7188	&	0.0002	&	20950.5	&	-0.0192	&	2459961.8775	&	0.0002	&	20985.5	&	-0.0223	\\
2459947.9169	&	0.0002	&	20951	&	-0.0234	&	2459962.0808	&	0.0001	&	20986	&	-0.0213	\\
2459948.1230	&	0.0002	&	20951.5	&	-0.0196	&	2459962.2822	&	0.0001	&	20986.5	&	-0.0223	\\
2459948.3216	&	0.0002	&	20952	&	-0.0233	&		&		&		&		\\
\hline
\hline
\end{tabular}
\end{center}
\label{tabA2}
\end{table*}

\clearpage
\bibliography{Ref}{}
\bibliographystyle{aasjournal}
\end{document}